# Interferometric fiber-optic bending / nano-displacement sensor using plastic dual-core fiber

H. Qu,[1] G. F. Yan,[1,2] and M. Skorobogatiy[1,*]

[1]*Genie Physique, Ecole Polytechnique de Montreal, C. P. 6079. Succ. Centre-ville, Montreal, Quebec, Canada, H3C 3A7*
[2]*Center for Optical and Electromagnetic Research, Zhejiang University, Hangzhou, China 310058*
*\*Corresponding author: hang.qu@polymtl.ca*

We demonstrate an interferometric fiber-optic bending/micro-displacement sensor based on a plastic dual-core fiber with one end coated with a silver mirror. The two fiber cores are first excited with the same laser beam, the light in each core is then back-reflected at the mirror-coated fiber-end, and, finally, the light from the two cores is made to interfere at the coupling end. Bending of the fiber leads to shifting interference fringes that can be interrogated with a slit and a single photodetector. We find experimentally that the resolution of our bending sensor is ~$3\times10^{-4}$ m$^{-1}$ for sensing of bending curvature, as well as ~70 nm for sensing of displacement of the fiber tip. We demonstrate operation of our sensor using two examples. One is weighting of the individual micro-crystals of salt, while the other one is monitoring dynamics of isopropanol evaporation.

The R&D into low-cost and highly sensitive fiber-optic bending / displacement sensors is an active research field stimulated by large demand in various scientific and industrial sectors such as bio- and physiological sensing, architecture, robotics, astronautics and automotive industry [1]. To date, various types of fiber-optic bending sensors have been proposed. Bending sensors using a single piece of regular single-mode fiber or multimode fiber constitute the simplest type of a fiber bending sensor [2-4]. These sensors typically operate using intensity-based detection modality according to which the fiber transmitted intensity is characterized as a function of the fiber bending curvature. Among the advantages of these sensors are low cost, ease of fabrication, and simple signal acquisition and processing.. However, the detection accuracy of these intensity-based sensors is prone to errors due to intensity fluctuations of the light source, as well as due to lack of a reference channel that is desirable in order to increase tolerance to environmental variations. The bending-curvature resolution of these sensors is typically 0.01-0.5 m$^{-1}$ [2-4].

Another important type of a fiber bending sensor is based on fiber gratings that include fiber Bragg gratings and long period fiber gratings [5-10]. Bending of a fiber grating leads to variations in the pitch and refractive index (photoelastic effect) of the grating, thus shifting its resonant wavelengths. Compared to simple intensity-based sensors, fiber grating-based sensors are more sensitive, and they allow simultaneous sensing of several measurands such as temperature, strain and refractive index [5, 8]. At the same time, resolution and sensitivity of the fiber grating-based sensors is limited by the resolution (hence cost) of a spectrometer used in a setup. Assuming that 10 pm (typical resolution of a high-end optical spectrum analyzer) wavelength shift can be reliably detected, typical bending resolution of the fiber grating sensors is ~$10^{-3}$-$10^{-1}$ m$^{-1}$ [6-10]. It was also reported in [8] that spectral properties of the fiber gratings may degrade due to strong bending, thus leading to difficulties in finding resonant wavelengths.

Fiber bending sensors using in-fiber interferometry have been widely studied due to their very high sensitivities. In the interferometric fiber-optic bending sensors, the light is first split and guided along two different paths, or it is coupled into different optical modes. Recombination of light from the different paths or modes generates interference pattern that is sensitive to fiber bending. Most frequently, one uses spectrally broad light sources and a spectrometer to interrogate such sensors. Namely, one measures bending by interpreting changes in the spectral positions of the intensity maxima that are recorded at a fixed observation point. Alternatively, one can use monochromatic light source. In this case, fiber bending is quantified by detecting spatial shifts of the intensity fringes. Various implementations of the interferometric fiber-optic bending sensors have been proposed, including: splicing a multimode fiber or a photonic crystal fiber between two single-mode fibers [11-13], cascading two fiber-tapers [14], using multiple fiber gratings [15], or using multi-core fibers [16-19]. Sensitivities of the interferometric fiber-optic bending sensors operating in spectral interrogation mode can achieve 36 nm/m$^{-1}$ [15], which is equivalent to resolution of bending curvature ~$3\times10^{-4}$ m$^{-1}$, assuming that a spectral shift of 10 pm could be reliably detected. Majority of the interferometric fiber sensors use heterogeneous fiber structures and require complex splicing, which makes their fabrication difficult and costly, while also reflecting negatively on the sensor reliability. Similarly to the fiber grating-based sensors, resolution of the interferometric fiber sensors is frequently limited by the resolution of the spectrometers used for spectral interrogation.

In this paper, we demonstrate an interferometric fiber-optic bending/nano-displacement sensor based on a simple plastic dual-core fiber. Such fibers can be very cheaply fabricated using stack-and-draw technique with commodity plastics (see Fig. 1). Moreover, we use amplitude interrogation with a single photodetector and a slit, thus avoiding using costly spectrum analyzers, while demonstrating outstanding sensitivities comparable to the best ones reported for the interferometric fiber sensors. In our sensor, one end of the dual-core fiber is wet-coated with a silver mirror [20]. Thus, the light launched into the two cores is first guided along the fiber, and then reflected by the mirror at the fiber end. Reflected light comes out of the fiber coupling end and forms an interferogram (see Fig. 2). For signal detection we use a single photodetector placed behind a narrow slit with a size comparable to the size of an individual intensity fringe. Among the advantages of the dual-core fiber sensor are low cost, simple sensor structures, ease of fabrication, high sensitivity, and simplicity of sensing mechanism. We demonstrate operation of our sensor on the examples of weighting individual micro-crystals of salt with resolution of ~0.2 mg, and studying dynamics of isopropanol evaporation with evaporation rate of ~0.016 mg/s.

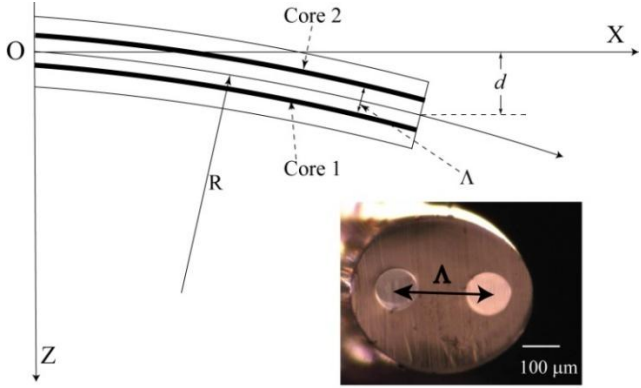

Fig.1 Schematic of the dual-core fiber bent in the X-Z plane. In the insert we show cross section of the dual-core fiber with one of its cores lit up. R is the bending curvature of the fiber; $\Lambda$ is the distance between the two cores; $d$ is the displacement of the fiber.

The dual-core fiber used in the sensor is fabricated in-house using a standard plastic fiber drawing tower. The fiber perform is fabricated by first drilling two parallel channels (~6.3 mm diameter each) in a polymethyl methacrylate (PMMA) rod of 25.6 mm diameter. Then, two polycarbonate (PC) rods are inserted into the channels to form the fiber preform. The preform is then consolidated for 1h in the vacuum oven at 120 ℃, and then drawn into fiber of ~ 550 μm outer diameter at 185 ℃. All the materials used in preform fabrication are commodity plastics purchased from McMaster Carr. The resultant fiber (see Fig. 1) comprises two polycarbonate cores (n~1.59) surrounded by PMMA cladding (n~1.48). The diameter of an individual core is ~110 μm, thus resulting in a multimode guidance. In Fig. 1, we present a sketch of the bent dual-core fiber, as well as a photograph of the fiber crossection. The typical fiber loss of a dual core fiber at ~630 nm is estimated to be 25 dB/m. Before building a dual-core fiber sensor, we also verify that the crosstalk between the two fiber cores over 10 cm propagation is negligible (less than -42 dB). To do this measurement we couple laser beam into one of the fiber cores and then use CCD camera to record the power in the other fiber core.

When bending the fiber in the X-Z plane, a differential strain between the two cores is introduced, thus leading to variations both in the refractive index (photoelastic effect) and the lengths of the two cores. When the fiber is displaced by $d$, changes in the phase difference ($\delta\phi$) of the light propagating in the two cores can be calculated as:

$$\delta\phi \approx 2k_0[(l\delta n_1 + n\delta l_1) - (l\delta n_2 + n\delta l_2)]$$
$$\approx 2k_0 nl[(\delta n_1 - \delta n_2)/n + (\delta l_1 - \delta l_2)/l] \quad , \quad (1)$$

where $\delta l_1$ and $\delta l_2$ are the changes in the lengths of the cores 1 and 2; $\delta n_1$ and $\delta n_2$ are the corresponding changes in the core refractive indices; $n$ and $l$ are the refractive index and the length of the two cores in a straight fiber; $k_0$ is the wavenumber defined as $2\pi/\lambda$, where $\lambda$ is wavelength. The difference in strain between the two cores can be simply associated with the bending radius as:

$$(\delta l_1 - \delta l_2)/l = \Lambda/R \quad , \quad (2)$$

where $\Lambda$ is the distance between the two cores, and R is the fiber bending radius. Moreover, due to the photoelastic effect, fiber bending also results in changes in the refractive index of the two cores. According to [21], the bending-induced variation in the refractive index of the two cores can be expressed as:

$$(\delta n_1 - \delta n_2)/n = C\Lambda n^2/R \quad , \quad (3)$$

where $C$ is a constant determined by the photoelastic tensors of the core material. Therefore, by substituting Eq. (2) and (3) into Eq. (1), we obtain:

$$\delta\phi = 4\pi n(1+Cn^2)\cdot(l/R)\cdot(\Lambda/\lambda) \quad . \quad (4)$$

Particularly, if the displacement $d$ of the fiber tip is much smaller than the fiber length $l$, the bending radius can be approximated as $R = l^2/2d$. In this case, form Eq. (4) we conclude that the phase difference is linearly proportional to the displacement of the fiber tip:

$$\delta\phi = 8\pi n(1+Cn^2)\cdot(d/l)\cdot(\Lambda/\lambda) \quad . \quad (5)$$

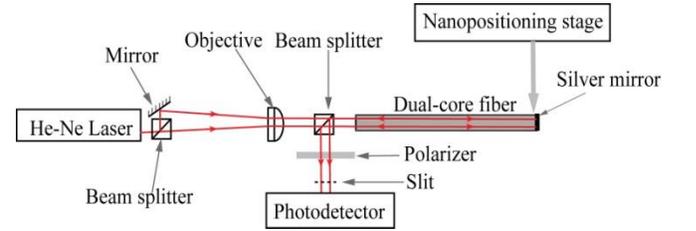

Fig.2. Schematic of the bending / nano-displacement sensor setup.

In Fig. 2, we present schematic of the experimental setup. A linearly-polarized He-Ne laser (wavelength: 632.8 nm) is split into two semi-parallel beams which pass through an objective and a beam splitter, and then are coupled into two fiber cores. Alternatively, one can avoid using objective and laser beam splitting altogether by simply using a metal mask deposited onto the coupling fiber facet with holes at the position of the two cores. One end of the dual-core fiber is immobilized in a V-groove, while the other end (free end) is wet-coated with a silver mirror. A glass plate fixed on a 3-D nanopositioning stage (Max341, Thorlabs) is used to displace the fiber tip. The total fiber length used in the setup is 30 mm, while the length of a bent portion of the fiber is $l$=15 mm. The light travels back and forth in the two cores and it finally comes out at the fiber coupling end. We then use a polarizer in front of a CCD camera in order to improve signal to noise ratio. The interferogram generated by the light beams coming out of the two fiber cores and reflected by the beam splitter can be recorded directly using a CCD camera (see Fig. 3). In order to reduce the cost of a sensor device we replace the CCD with a single photodetector placed behind a narrow slit with a size comparable to the size of a single interferometric fringe. When bending the fiber, the interference fringes shift due to variation in the phase difference between the two light beams coming from the fiber cores. In the multimedia attachment to this paper we show changes in the interferogram as fiber tip is displaced from 0 to 60 μm with a speed of 1 μm/s.

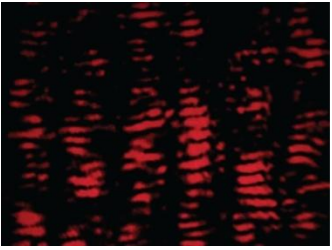

Fig. 3. (multimedia file) A typical interferogram generated by the two-core fiber-based bending sensor.

In order to extract the bending curvatures as a function of the fiber tip displacement (Fig. 4(b)) we analyse the fiber micrographs by fitting the fiber shapes with circular arcs (3-point-fitting). As seen from Fig. 4(b), even at zero displacement of the fiber tip, the curvature of the fiber can be non-zero. This is because the plastic fiber can have a natural curvature after drawing, or it can acquire some permanent curvature during storage in coils. By fixing the observation point, and by counting the number of fringes that pass the observation point in response to the fiber displacement (see Fig. 4(a)) we can calibrate the dual-core fiber sensor for bending/nano-displacement sensing.

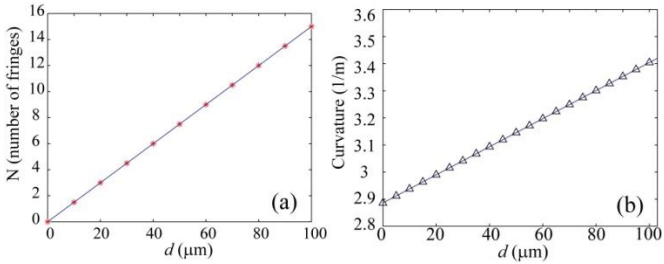

Fig. 4. (a) Number of fringes that pass observation point during fiber tip displacement. (b) Bending curvature of the fiber as a function of the fiber tip displacement.

Fig. 4 suggests that the number of counted fringes ($N = \delta\phi / 2\pi$) is linearly proportional to the displacement of the fiber tip, as well as to the fiber curvature, which is in accordance with Eqs. (4, 5). Form this data we can calculate sensitivity of the dual-core fiber sensor, which is found to be 0.15 fringe/μm for displacement sensing and 30 fringe/m$^{-1}$ for bending curvature sensing. Assuming a primitive electronic circuit that resolves only a change from the maximum to the minimum of the fringe intensity (0.5 fringe shift) the detection limit of even the simplest interferometric sensor will be ~3 μm (tip shift) and ~0.02 m$^{-1}$ (bending curvature). In fact, almost two orders of sensitivity improvement can be gained by analyzing full intensity curves as shown in what follows.

In a practical bending sensor, instead of an expensive CCD camera one would rather use a single photodetector that samples a small area of the interferogram. In our setup we use a 918D Low-Power Photodiode Sensor (Newport) that is placed behind a 1mm-long, 125 μm-wide slit. The integration time constant of a photodiode is 1ms. In Fig. 5 (a-c), we present intensity variations measured by the photodiode, while the fiber is displaced by 20 μm, 40 μm, and 60 μm with a speed of 1 μm/s. The intensity oscillations in Fig. 5 are a direct consequence of the fringe shifts caused by the fiber bending. Therefore, fiber bending can be quantified simply by counting the number of the periods in the intensity variation. Note also a good repeatability of the sensor response. Particularly, Figs. 5 (a, b) look simply like cutouts of Fig. 5(c), although the three measurements are completely independent.

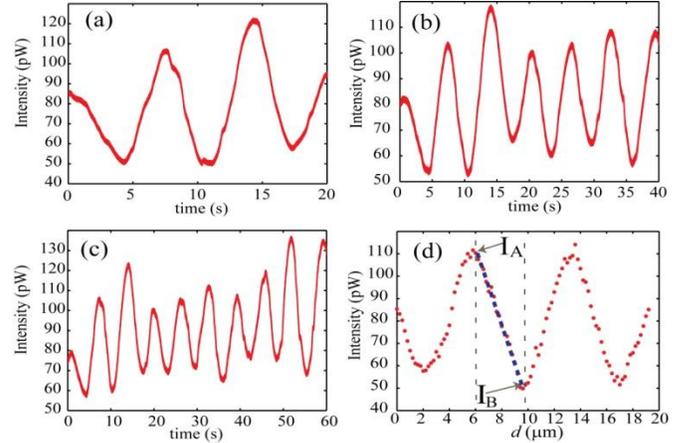

Fig. 5. Intensity variations measured by a photodiode placed behind a slit. Three distinct measurements corresponding to (a) 20 μm, (b) 40 μm, and (c) 60 μm displacements of the fiber tip with a speed of 1 μm/s. (d) Intensity variation while the fiber is step-by-step displaced from 0 to 20 μm with a 0.2 μm step.

As mentioned earlier, analysis of intensity changes between the two fringes, rather than just counting the number of fringes, may provide a much higher sensitivity. In Fig 5(d) we present recorded intensity variation while the fiber tip is displaced step-by-step from 0 to 20 μm with a 0.2 μm interval. Each point in Fig. 5(d) represents an average value of the intensity recorded using 1ms integration time constant of photodetector and 20s-long acquisition time. Note that while the measurements in Figs. 5(a) and 5(d) use different interrogation modalities for sensing 20μm displacement of the fiber tip (continuous displacement versus step-by-step displacement), the sensor response shows good repeatability in both measurements. Moreover, in Fig. 5(d), registered intensity depends linearly on the fiber tip displacement between the fringe maxima and minima.. Consider, for example, the second falling edge that corresponds to the tip displacement in the range 6-10 μm. Here, the difference between the fringe maximum ($I_A$) and minimum ($I_B$) is ~60 pW, with a corresponding power sensitivity to displacement being ~15 pW/μm. Considering that the responsivity of our photodiode detector to intensity variation is ~1 pW, we estimate that the detection limit of our fiber sensor in this range can achieve ~0.07 μm for the displacement sensing and ~3×10$^{-4}$ m$^{-1}$ for the bending curvature sensing.

Finally, we present two examples of using our sensor to perform static and dynamic measurements. In the first example, we use our sensor to measure the weight of salt particles (static measurement), while in the second example we measure the changing weight of the isopropanol droplets while solvent evaporates (dynamic measurement). In the static experiment with salt crystals, we calibrate our sensor by adding micro-crystals of table salt one by one into a small sample holder attached at the tip of a double-core fiber. The individual crystals are of ~2 mg average weight and they are weighted using electronic balance before being placed into the sample holder. The recorded intensity is presented in Fig. 6 (a). As the net weight of salt particles increases, the measured intensity shows periodical oscillations. From the slope of intensity curve between the fringe maxima and minima we conclude that the weight change of 1 mg results in the fiber tip displacement of ~0.3 μm. Assuming ~1 pW detector responsivity, the resolution of our fiber sensor for weighting and force sensing is estimated to be ~0.2 mg (equivalent to ~$2\times10^{-3}$ N).

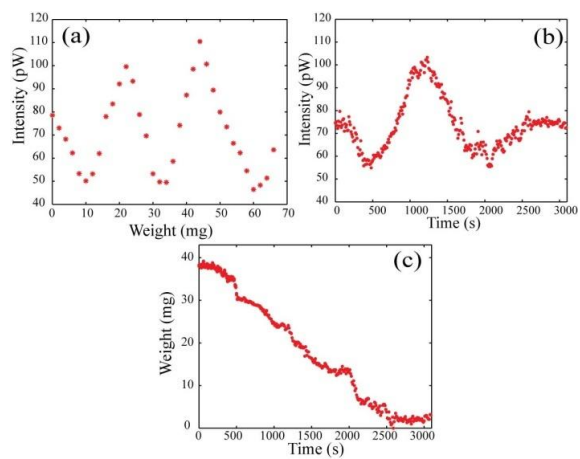

Fig. 6. (a) Intensity variations as a function of the weight of salt particles. (b) Intensity variations as isopropanol evaporates from a textile thread. (c) changes in weight as isopropanol evaporates.

In the second experiment, we add two drops of isopropanol to the sample holder, and detect variations in the measured intensity versus time as isopropanol evaporates. Each intensity point in Fig. 6(b) is an average over 10 s acquisition time. Comparing Figs. 6(a) and 6(b) allows us to extract the average solvent evaporation rate. For example, solvent weight loss during the time period between 500-2000s corresponds to the 24 mg weight loss, which is judged from a full passage of a single fringe across the detector. This defines the evaporation rate of ~0.016 mg/s. Moreover, we can use Fig. 6(a) as a calibration curve in order to extract from Fig. 6(b) the dynamics of weight change of isopropanol. Remembering that the fiber movement in the evaporation experiment is in the opposite direction than in the salt weighting experiment, in Fig. 6(c) we present the final result of our paper, which is a time dependent change in the weight of evaporating isopropanol.

In conclusion, we demonstrate an interferometric bending/nano-displacement sensor based on a plastic dual-core fiber with one end coated with a silver mirror. Bending of the fiber leads to variations in the phase difference between the reflected light coming out of the two fiber cores, thus leading to shifts in the interference fringes. The fringe shifts can be then interrogated using a single photodetector placed behind a narrow slit. When used in force sensing or weight measurements, plastic dual-core fibers bring an additional advantage over silica glass-based fibers. Particularly, as the Young's modules of PMMA and polycarbonate (1.8-3 GPa) are much smaller than that of silica fibers (~70 GPa) [22], the same force would lead to considerably larger displacements of the plastic fibers as compared to glass fibers. Experimentally, the resolution of the fiber sensor is found to be ~$3\times10^{-4}$ m$^{-1}$ for bending curvature sensing and ~70 nm for displacement sensing, which are comparable to that of the most sensitive interferometric fiber-optic sensors. The advantages of our sensors include low cost, high sensitivity, simple sensor structure, ease of fabrication and simplicity of sensing mechanism.